\documentclass[11pt]{article}
\usepackage{parskip}
\usepackage{harvard}
\usepackage[utf8]{inputenc}
\usepackage{listings}
\usepackage{amsmath}
\usepackage{wasysym}
\usepackage{hyperref}
\usepackage{tabularx}

\usepackage{amsfonts}
\usepackage{amssymb}
\usepackage{amsthm}
\usepackage{multirow}
\usepackage{threeparttable}
\usepackage{graphicx}
\usepackage{epstopdf}
\usepackage[a4paper, total={6.5in,10in}]{geometry}
\usepackage{caption}
\usepackage{authblk}

\usepackage[titletoc,title]{appendix}
\usepackage{proof-at-the-end}

\newtheorem*{thm*}{Theorem}

\newtheorem*{lemma*}{Lemma}

\newcommand{\btheta}{\ensuremath{\boldsymbol\theta}}

\newcommand{\bx}{\ensuremath{\mathbf{x}}}

\newcommand{\bone}{\ensuremath{\mathbf{1}}}

\citationstyle{agsm}
\title{Reference and Probability-Matching Priors for the Parameters of a Univariate Student $t$-Distribution}
\author{A. J. van der Merwe}
\author{M. J. von Maltitz}
\affil{Department of Mathematical Statistics and Actuarial Science, University of the Free State, Bloemfontein, SA}
\author{J. H. Meyer}
\affil{Department of Mathematics and Applied Mathematics, University of the Free State, Bloemfontein, SA}
\date{12 November 2020}
\begin{document}

\appendixtitleon
\appendixtitletocon

\maketitle

\begin{abstract}
In this paper reference and probability-matching priors are derived for the univariate Student $t$-distribution. These priors generally lead to procedures with properties frequentists can relate to while still retaining Bayes validity. The priors are tested by performing simulation studies. The focus is on the relative mean squared error from the posterior median ($MSE(\nu)/\nu$) and on the frequentist coverage of the 95\% credibility intervals for a sample size of $n=30$. Average interval lengths of the credibility intervals as well as the modes of the interval lengths based on 2000 simulations are also considered. The performance of the priors are also tested on real data, namely daily logarithmic returns of IBM stocks. 

\vspace{\baselineskip}
\noindent \textbf{Keywords.} Reference priors; Probability-matching prior; $t$-Distribution; Mean squared error; Credibility intervals; Coverage percentages; Log-returns 
\end{abstract}

\section{Introduction} \label{sec:intro}
In most applied as well as theoretical research works, the residual terms in linear models are assumed to be normally and independently distributed. However, such assumptions may not be appropriate in many practical situations (see for example \citeasnoun{gnan72} and \citeasnoun{zellner76}). Many economic and business data, for example stock return data, exhibit heavy (or fat) tail distributions and cannot be effectively modelled by the normal distribution. The use of the Student $t$-distribution reduces the influence of outliers and thus makes the statistical analysis more robust \cite{fonseca08}. The smaller the number of degrees of freedom, the more robust the analysis will be. The suitability of the $t$-distribution to model outliers has been thoroughly discussed in the literature and has been applied in disciplines such as stock return data \cite{blatt74,zellner76}, medicine \cite{liu97}, global navigation satellite systems \cite{vaneck96}, finance and biology \cite{fernandez98} and portfolio optimisation \cite{kotz04}.

Unfortunately, the estimation of $\nu$, the number of degrees of freedom of the $t$-distribution, is not easy. The reason for this is the bad behaviour of the likelihood function for $\nu$ for a given location and scale parameter. The likelihood function does not always go to zero if $\nu$ goes to infinity but tends to a positive constant. To overcome the fact that the likelihood function does not vanish in the tail a prior distribution that tends to zero as $\nu$ tends to infinity should be used to form a proper posterior distribution. The uniform prior will result in an improper posterior distribution for $\nu$ and can therefore not be used. It is for this reason that non-informative priors are derived in this paper. For further discussion on proper and improper priors for $\nu$ \citeaffixed{fonseca08,villa14}{see for example}. 

The manuscript is organized as follows. In Section~\ref{sec:priors} reference and probability-matching priors are given for the parameters $\nu$, $\mu$ and $\sigma^2$ of the univariate $t$-distribution. The proofs of these priors are given in Appendix A and in Appendix B it is shown that the priors tend to zero as $\nu$ tends to infinity, and that the reference priors result in proper posterior distributions. In Section 3 simulation studies are performed for standard $t$-distribution ($\mu=0$ and $\sigma^2=1$) based on the non-informative priors defined in Section 2 and on priors previously proposed. The focus is on the relative square-rooted mean squared error ($\sqrt{MSE(\nu)}/\nu$) from the posterior median and on the frequentist coverage of the 95\% credibility intervals for a sample of size $n=30$. Average interval lengths based on 2000 simulations are also considered. In Section 4 an application is given. 

\section{Reference and Probability-Matching Priors} \label{sec:priors}
Reference and probability-matching priors generally lead to procedures with properties frequentists can relate to while still retaining Bayesian validity. The derivation of the reference priors of \citeasnoun{berger92} depends on the ordering of the parameters and how the parameter vector is divided into sub-vectors. The reference prior maximizes the difference in information about the parameters provided by the prior and the posterior \cite{pearn05} \textit{i.e.} the reference prior provides as little information as possible about the parameters of interest. 

The probability-matching prior (another non-informative prior) on the other hand provides accurate frequentist intervals and is also used for comparisons in Bayesian analysis. \citeasnoun{datta95} provided a method for finding probability-matching priors by deriving a differential equation that a prior must satisfy if the posterior probability of a one-sided credibility interval for a parametric function and its frequentist probability agree up to $O\left(n^{-1}\right)$, where $n$ is the sample size. The following theorems can now be stated. 

\begin{theoremEnd}[category=A, end, restate, text link={\textbf{Proof:} See \hyperref[proof:prAtEnd\pratendcountercurrent]{proof} in Appendix \ref{proofsection:prAtEnd\pratendcountercurrent}}]{thm} \label{theorem:ref_prior}
The reference prior for the orderings $\{\nu,\mu, \sigma^2\}$, $\{\mu,\nu, \sigma^2\}$ and $\{\nu,\sigma^2, \mu\}$ is given by
\begin{align*}
    p_{2}\left(\nu,\mu, \sigma^2\right) & \propto \sigma^{-2} \left[\Psi'\left(\frac{\nu}{2}\right)-\Psi'\left(\frac{\nu+1}{2}\right)-\frac{2\left(\nu+3\right)}{\nu\left(\nu+1\right)^{2}} \right]^{\frac{1}{2}} 
\end{align*}
and the reference prior for the orderings $\{\mu,\sigma^2, \nu\}$, $\{\sigma^2, \mu, \nu\}$ and $\{\sigma^2, \nu, \mu\}$ is given by
\begin{align*}
    p_{1}\left(\mu,\sigma^2,\nu\right) & \propto \sigma^{-2} \left[\Psi'\left(\frac{\nu}{2}\right)-\Psi'\left(\frac{\nu+1}{2}\right)-\frac{2\left(\nu+5\right)}{\nu\left(\nu+1\right)\left(\nu+3\right)}\right]^{\frac{1}{2}} 
\end{align*}
where $\Psi\left(a\right) = \frac{d}{da}\log \Gamma\left(a\right)$ and $\Psi'\left(a\right) = \frac{d}{da}\Psi\left(a\right)$, the trigamma function.
\end{theoremEnd}

\begin{proofEnd}
The Fisher information matrix for the ordering $\left\{\nu, \mu,\sigma\right\}$ is given in \citeasnoun{fonseca08}, as,
\begin{align} \label{eq:fi}
I\left(\nu, \mu,\sigma \right) & = \left[ \begin{array}{ccc} \frac{n}{4}\left[\Psi'\left(\frac{\nu}{2}\right)-\Psi'\left(\frac{\nu+1}{2}\right)- \frac{2\left(\nu+5\right)}{\nu\left(\nu+1\right)\left(\nu+3\right)}  \right] & 0 & \frac{-2n}{\sigma\left(\nu+1\right)\left(\nu+3\right)}  \\ & & \\0 & \frac{n\left(\nu+1\right)}{\sigma^{2}\left(\nu+3\right)} & 0 \\ & & \\  \frac{-2n}{\sigma\left(\nu+1\right)\left(\nu+3\right)} & 0 & \frac{2n\nu}{\sigma^{2}\left(\nu+3\right)} \end{array} \right]  \\
 & = \left[ \begin{array}{ccc} F_{11} & F_{12} & F_{13}  \\ F_{21} & F_{22} & F_{23} \\ F_{31} & F_{32} & F_{33} \end{array} \right]  \nonumber
\end{align}
To calculate the reference prior for the ordering $\left\{\nu,\mu,\sigma \right\}$, we must first calculate:
\begin{align*}
h_{1} & = F_{11} - \left[\begin{array}{cc} F_{12} & F_{13} \end{array}\right]\left[ \begin{array}[c]{cc} F_{22} & F_{23} \\ F_{32} & F_{33} \end{array}\right]^{-1} \left[ \begin{array}{c} F_{21} \\ F_{31} \end{array} \right]  
\end{align*}
from the information matrix in Equation~(\ref{eq:fi}). Therefore,

\begin{align*}
h_{1} & = \frac{n}{4}\left[\Psi'\left(\frac{\nu}{2}\right)-\Psi'\left(\frac{\nu+1}{2}\right)- \frac{2\left(\nu+5\right)}{\nu\left(\nu+1\right)\left(\nu+3\right)}  \right] \\
& \hspace{1em}- \left[\begin{array}{cc} 0 &  \frac{-2n}{\sigma\left(\nu+1\right)\left(\nu+3\right)} \end{array}\right]\left[ \begin{array}[c]{cc}  \frac{\sigma^{2}\left(\nu+3\right)}{n\left(\nu+1\right)} & 0 \\ 0 & \frac{\sigma^{2}\left(\nu+3\right)}{2n\nu} \end{array}\right] \left[ \begin{array}{c}0 \\ \frac{-2n}{\sigma\left(\nu+1\right)\left(\nu+3\right)}  \end{array} \right]  \\
& =  \frac{n}{4}\left[\Psi'\left(\frac{\nu}{2}\right)-\Psi'\left(\frac{\nu+1}{2}\right) - \frac{2\left(\nu+3\right)}{\nu\left(\nu+1\right)^{2}}\right]
\end{align*}
Therefore,
\begin{align*}
h_{1}^{\frac{1}{2}} & \propto \left[\Psi'\left(\frac{\nu}{2}\right)-\Psi'\left(\frac{\nu+1}{2}\right) - \frac{2\left(\nu+3\right)}{\nu\left(\nu+1\right)^{2}}\right]^{\frac{1}{2}} 
\end{align*}

Now $p\left(\nu\right) \propto h_{1}^{\frac{1}{2}}$. Also, since it does not contain $\mu$, $h_{2}^{\frac{1}{2}}\propto c$. So $p\left(\mu | \nu\right) \propto c$. Further, $h_{3} = F_{33}=\frac{2n\nu}{\sigma^{2}\left(\nu+3\right)}$ and $p\left(\sigma | \nu,\mu\right) \propto h_{3}^{\frac{1}{2}} = \sigma^{-1}$. Therefore, 

\begin{align*}
p_{2}\left(\nu,\mu,\sigma\right) &= p\left(\nu\right) p\left(\mu|\nu\right) p\left(\sigma|\nu,\mu\right) \nonumber \\
& \propto \sigma^{-1}\left[\Psi'\left(\frac{\nu}{2}\right)-\Psi'\left(\frac{\nu+1}{2}\right) - \frac{2\left(\nu+3\right)}{\nu\left(\nu+1\right)^{2}}\right]^{\frac{1}{2}}.
\end{align*}

Similarly, $p_{2}\left(\nu,\mu,\sigma^{2}\right) \propto \sigma^{-2}\left[\Psi'\left(\frac{\nu}{2}\right)-\Psi'\left(\frac{\nu+1}{2}\right) - \frac{2\left(\nu+3\right)}{\nu\left(\nu+1\right)^{2}}\right]^{\frac{1}{2}}$ because it is known that if $p\left(\sigma\right) \propto \sigma^{-1}$, then $p\left(\sigma^{2}\right) \propto \sigma^{-2}$, and for $p\left(\sigma\right) \propto \sigma^{-2}$ it follows that $p\left(\sigma^{2}\right) \propto \sigma^{-3}$.

By using the Fisher information matrices $I\left(\nu,\sigma,\mu\right)$ and $I\left(\mu,\nu,\sigma \right)$ is can be shown that $p_{2}\left(\nu,\mu,\sigma\right)$ is also a reference prior for the orderings $\left\{\nu,\sigma,\mu\right\}$ and $\left\{\mu,\nu,\sigma\right\}$. In a similar way it can be proved that $p_{1}\left(\mu,\sigma,\nu\right) \propto \sigma^{-1} \left[\Psi'\left(\frac{\nu}{2}\right)-\Psi'\left(\frac{\nu+1}{2}\right)-\frac{2\left(\nu+5\right)}{\nu\left(\nu+1\right)\left(\nu+3\right)}\right]^{\frac{1}{2}}$ is a reference prior for the orderings $\left\{\mu,\sigma,\nu\right\}$, $\left\{\sigma,\mu,\nu\right\}$ and $\left\{\sigma,\nu,\mu\right\}$. For further details on reference priors see \citeasnoun{berger92}.

\end{proofEnd}

\begin{theoremEnd}[category=A, end, restate, text link={\textbf{Proof:} See \hyperref[proof:prAtEnd\pratendcountercurrent]{proof} in Appendix \ref{proofsection:prAtEnd\pratendcountercurrent}}]{thm} \label{theorem:pm_prior}
$p_{2}\left(\nu,\mu, \sigma^2\right)$ is also a probability-matching prior for $\nu$.
\end{theoremEnd}

\begin{proofEnd}
To derive the probability-matching prior $P_{M}\left(\nu, \mu, \sigma\right)$, we need the inverse of the Fisher information matrix, 
\begin{align*}
I^{-1}\left(\mu,\sigma,\nu \right) &= \left[ \begin{array}{ccc} \frac{\sigma^{2}\left(\nu+3\right)}{n\left(\nu+1\right)} & 0 & 0 \\ & & \\ 0 & \frac{n}{4D}\left[\Psi'\left(\frac{\nu}{2}\right)-\Psi'\left(\frac{\nu+1}{2}\right)- \frac{2\left(\nu+5\right)}{\nu\left(\nu+1\right)\left(\nu+3\right)}\right] & \frac{2n}{D\sigma\left(\nu+1\right)\left(\nu+3\right)} \\ & & \\ 0 & \frac{2n}{D\sigma\left(\nu+1\right)\left(\nu+3\right)} & \frac{2n\nu}{D\sigma^{2}\left(\nu+3\right)} \end{array} \right] \nonumber
\end{align*}
where 
\begin{align*}
D &= \frac{n^{2}\nu}{2\sigma^{2}\left(\nu+3\right)}\left[\Psi'\left(\frac{\nu}{2}\right)-\Psi'\left(\frac{\nu+1}{2}\right)- \frac{2\left(\nu+5\right)}{\nu\left(\nu+1\right)\left(\nu+3\right)}\right] - \frac{4n^{2}}{\sigma^{2}\left(\nu+1\right)^{2}\left(\nu+3\right)^{2}}  \\
& = \frac{n^{2}\nu}{2\sigma^{2}\left(\nu+3\right)}\left[\Psi'\left(\frac{\nu}{2}\right)-\Psi'\left(\frac{\nu+1}{2}\right) - \frac{2\left(\nu+3\right)}{\nu\left(\nu+1\right)^{2}}\right] 
\end{align*}

Let $t\left(\btheta\right) = \nu$, where  $t\left(\btheta\right)$ is the parameter of interest. 

From this it follows that $\frac{\partial t\left(\btheta\right)}{\partial\nu} = 1; \frac{\partial t\left(\btheta\right)}{\partial\mu} = 0; \frac{\partial t\left(\btheta\right)}{\partial\sigma} = 0$, and,
\begin{align*}
\nabla_{t}'\left(\btheta\right) & = \left[\begin{array}{ccc} \frac{\partial t\left(\btheta\right)}{\partial\mu} & \frac{\partial t\left(\btheta\right)}{\partial\sigma} & \frac{\partial t\left(\btheta\right)}{\partial\nu} \end{array} \right] \\
& = \left[\begin{array}{ccc} 0 & 0 & 1 \end{array}\right]
\end{align*}

Therefore,
\begin{align*}
\nabla_{t}'\left(\btheta\right)I^{-1}\left(\btheta \right) & = \left[\begin{array}{ccc} 0 & \frac{2n}{D\sigma\left(\nu+1\right)\left(\nu+3\right)} & \frac{2n\nu}{D\sigma^{2}\left(\nu+3\right)} \end{array}\right] ,
\end{align*}
which means that,
\begin{align*}
\left[\nabla_{t}'\left(\btheta\right)I^{-1}\left(\btheta \right)\nabla_{t}\left(\btheta\right)\right]^{\frac{1}{2}} & = \left(\frac{2n\nu}{D\sigma^{2}\left(\nu+3\right)}\right)^{\frac{1}{2}}. 
\end{align*}
\begin{align*}
\zeta'\left(\btheta\right) & = \frac{\nabla_{t}'\left(\btheta\right)I^{-1}\left(\btheta \right)}{\left[\nabla_{t}'\left(\btheta\right)I^{-1}\left(\btheta \right)\nabla_{t}\left(\btheta\right)\right]^{\frac{1}{2}}} \\
& = \left[\begin{array}{ccc} \zeta_{1}\left(\btheta\right) & \zeta_{2}\left(\btheta\right) & \zeta_{3}\left(\btheta\right) \end{array}\right] \\
& = \left[\begin{array}{ccc} 0 & \frac{\left(2n\right)^{\frac{1}{2}}}{D^{\frac{1}{2}}\nu^{\frac{1}{2}}\left(\nu+1\right)\left(\nu+3\right)^{\frac{1}{2}}} & \frac{\left(2n\nu\right)^{\frac{1}{2}}}{D^{\frac{1}{2}}\sigma\left(\nu+3\right)^{\frac{1}{2}}} \end{array}\right] \nonumber
\end{align*}
This indicates that the probability-matching prior is:
\begin{align*}
p_{M}\left(\btheta\right) = p_{M}\left(\nu, \mu, \sigma\right) & \propto D^{\frac{1}{2}} \frac{\left(\nu+3\right)^{\frac{1}{2}}}{\nu^{\frac{1}{2}}} \\
& \propto \sigma^{-1}\left[\Psi'\left(\frac{\nu}{2}\right)-\Psi'\left(\frac{\nu+1}{2}\right) - \frac{2\left(\nu+3\right)}{\nu\left(\nu+1\right)^{2}}\right]^{\frac{1}{2}}
\end{align*}
because the differential equation $\frac{\partial}{\partial \mu}\left[\zeta_{1}\left(\btheta\right) p\left(\btheta\right) \right] + \frac{\partial}{\partial \sigma}\left[\zeta_{2}\left(\btheta\right) p\left(\btheta\right) \right] + \frac{\partial}{\partial \nu}\left[\zeta_{3}\left(\btheta\right) p\left(\btheta\right) \right] = 0$. The probability-matching prior is therefore the same as the reference priors for the orderings $\left\{\nu,\mu,\sigma \right\}$, $\left\{\mu,\nu,\sigma \right\}$, and $\left\{\nu,\sigma,\mu \right\}$.
\end{proofEnd}

\begin{theoremEnd}[category=B, end, restate, text link={\textbf{Proof: }See \hyperref[proof:prAtEnd\pratendcountercurrent]{proof} in Appendix \ref{proofsection:prAtEnd\pratendcountercurrent}}]{thm}  \label{theorem:tend_zero}
The reference priors tend to zero as $\nu$ tends to infinity. 
\end{theoremEnd}

\begin{proofEnd}
The proof is the same as that of the Corollary 1 in \citeasnoun{fonseca08}. Consider $\left[p_{2}\left(\nu\right)\right]^{2} = \left[\Psi'\left(\frac{\nu}{2}\right)-\Psi'\left(\frac{\nu+1}{2}\right)-\frac{2\left(\nu+3\right)}{\nu\left(\nu+1\right)^{2}} \right]$. By using Stirling's asymptotic formula $\Psi'\left(a\right) \approx a^{-1}+\left(2a^{2}\right)^{-1}$, for large $a$ it follows that,
\begin{align*}
    \Psi'\left(\frac{\nu}{2}\right) \approx \left(\frac{\nu}{2}\right)^{-1}+\left[2\left(\frac{\nu}{2}\right)^{2}\right]^{-1} = \frac{2}{\nu} + \frac{2}{\nu^{2}}
\end{align*}
and,
\begin{align*}
    \Psi'\left(\frac{\nu+1}{2}\right) \approx \frac{2}{\nu+1} + \frac{2}{\left(\nu+1\right)^{2}}.
\end{align*}
Therefore, 
\begin{align*}
    \Psi'\left(\frac{\nu}{2}\right)- \Psi'\left(\frac{\nu+1}{2}\right) &\approx \frac{2\nu^{2}+6\nu+2}{\nu^{2}\left(\nu+1\right)^{2}},
\end{align*}
and,
\begin{align*} 
\left[p_{2}\left(\nu\right)\right]^{2}&= \frac{2\nu^{2}+6\nu+2}{\nu^{2}\left(\nu+1\right)^{2}} -\frac{2\left(\nu+3\right)}{\nu\left(\nu+1\right)^{2}} \\
    &= \frac{2}{\nu^{2}\left(\nu+1\right)^{2}} \\
    &= O\left(\nu^{-4}\right)
\end{align*}
Therefore $p_{2}\left(\nu\right)=O\left(\nu^{-2}\right) \textnormal{ as } \nu \to \infty$.

In a similar way it can be proved that $\left[p_{1}\left(\nu\right)\right]^{2} = \frac{2\left(5\nu+3\right)}{\nu^{2}\left(\nu+1\right)^{2}\left(\nu+3\right)}$, which means that $p_{1}\left(\nu\right)=O\left(\nu^{-2}\right) \textnormal{ as } \nu \to \infty$.
\end{proofEnd}

\begin{theoremEnd}[category=B, end, restate, text link={\textbf{Proof: }See \hyperref[proof:prAtEnd\pratendcountercurrent]{proof} in Appendix \ref{proofsection:prAtEnd\pratendcountercurrent}}]{thm}  \label{theorem:proper}
In the case of the standard univariate $t$-distribution the reference priors result in proper posterior distributions for $\nu$.
\end{theoremEnd}

\begin{proofEnd}
The proof will be given for $p_{1}\left(\nu\right)$. The proof for $p_{2}\left(\nu\right)$ follows in a similar way. The posterior for $\nu$ is as follows:
\begin{align*}
    p_{1}\left(\nu|data \right) & = \tilde{k}\left[\Psi'\left(\frac{\nu}{2}\right)-\Psi'\left(\frac{\nu+1}{2}\right)-\frac{2\left(\nu+5\right)}{\nu\left(\nu+1\right)\left(\nu+3\right)}\right]^{\frac{1}{2}} \\
    & \hspace{2em} \times \frac{\Gamma\left(\frac{\nu+1}{2}\right)^{n}\nu^{n\nu/2}}{\Gamma\left(\frac{\nu}{2}\right)^{n}\Gamma\left(\frac{1}{2}\right)^{n}}\left[\prod_{i=1}^{n}\left(\nu+x_{i}^{2}\right)\right]^{-\frac{1}{2}\left(\nu+1\right)},
\end{align*}
where $\tilde{k}$ is the normalizing constant. We then have that:
\begin{align*}
    p_{1}\left(\nu|data \right) & \leq \tilde{k}\left[\Psi'\left(\frac{\nu}{2}\right)\right]^{\frac{1}{2}} \times \frac{\Gamma\left(\frac{\nu+1}{2}\right)^{n}\nu^{n\nu/2}}{\Gamma\left(\frac{\nu}{2}\right)^{n}\left(\sqrt{\pi}\right)^{n}}\left[\prod_{i=1}^{n}\left(\nu+x_{i}^{2}\right)\right]^{-\frac{1}{2}\left(\nu+1\right)}.
\end{align*}
Since $\left(\nu^{\nu}\right)^{n/2} \to 1^{n/2} = 1$ if $\nu \to 0^{+}$, it follows that, if $\nu \to 0^{+}$, then
\begin{align*}
    \left[\prod_{i=1}^{n}\left(\nu+x_{i}^{2}\right)\right]^{-\frac{1}{2}\left(\nu+1\right)} \to \left[\prod_{i=1}^{n}x_{i}^{2}\right]^{-\frac{1}{2}}
\end{align*}
It is therefore only necessary to consider: 
\begin{align*}
    \lim_{\nu \to 0^{+}} \frac{\left[\Psi'\left(\nu\right)\right]^{\frac{1}{2}}\Gamma\left(\nu+\frac{1}{2}\right)^{n}}{\Gamma\left(\nu\right)^{n}}.
\end{align*}
Since $\Psi\left(\nu\right) = \frac{d}{d\nu}\left[\ln \Gamma \left(\nu\right)\right] = \frac{\Gamma'\left(\nu\right)}{\Gamma\left(\nu\right)}$, it follows that,
\begin{align*}
    \Psi'\left(\nu\right) = \frac{\Gamma''\left(\nu\right)\Gamma\left(\nu\right)-\left[\Gamma'\left(\nu\right)\right]^{2}}{\left[\Gamma\left(\nu\right)\right]^{2}}.
\end{align*}
Therefore,
\begin{align}
    \lim_{\nu \to 0^{+}} \frac{\left[\Psi'\left(\nu\right)\right]^{\frac{1}{2}}\left[\Gamma\left(\nu+\frac{1}{2}\right)\right]^{n}}{\left[\Gamma\left(\nu\right)^{n}\right]} &= \lim_{\nu \to 0^{+}} \left\{\frac{\Gamma''\left(\nu\right)\Gamma\left(\nu\right)-\left[\Gamma'\left(\nu\right)\right]^{2}}{\left[\Gamma\left(\nu\right)\right]^{2}}\cdot \frac{\left[\Gamma\left(\nu+\frac{1}{2}\right)\right]^{2n}}{\left[\Gamma\left(\nu\right)\right]^{2n}}\right\}^{\frac{1}{2}}. \label{eq:2-4-1}
\end{align}
The following formulae are valid:
\begin{align}
    -\frac{\Gamma'\left(\nu\right)}{\Gamma\left(\nu\right)} = \frac{1}{\nu} + \gamma + \sum_{n=1}^{\infty}\left(\frac{1}{n+\nu}-\frac{1}{n}\right), \hspace{1em} \nu>0 \label{eq:2-4-2}
\end{align}
where $\gamma=0.5772$ is Euler's constant.
It can also be shown that if $\nu>0$ we have that
\begin{align*}
    \sum_{n=1}^{\infty}\left(\frac{1}{n+\nu}-\frac{1}{n}\right) = -\Psi\left(1+\nu\right)-\gamma,
\end{align*}
and that, if $\nu>0$,
\begin{align}
    \frac{1}{\Gamma\left(\nu\right)} = \nu \exp \left[\gamma\nu - \sum_{k=2}^{\infty} \frac{\left(-1\right)^{k}\zeta\left(k\right) \nu^{k}}{k}\right], \label{eq:2-4-3}
\end{align}
where $\zeta\left(k\right)$ is Riemann's zeta function. 
Therefore Equations~\ref{eq:2-4-2} $\times$ \ref{eq:2-4-3} gives
\begin{align*}
    -\frac{\Gamma'\left(\nu\right)}{\left[\Gamma\left(\nu\right)\right]^{2}} = \left[1+\nu\gamma-\nu\Psi\left(1+\nu\right)-\nu\gamma\right] \times \exp \left[\gamma\nu - \sum_{k=2}^{\infty} \frac{\left(-1\right)^{k}\zeta\left(k\right) \nu^{k}}{k}\right].
\end{align*}
Since $\Psi\left(1\right)=-\gamma$, it follows that, as $\nu \to 0^{+}$,
\begin{align}
    -\frac{\Gamma'\left(\nu\right)}{\left[\Gamma\left(\nu\right)\right]^{2}} &\to 1. \nonumber
\end{align}    
Therefore,
\begin{align}    
    \frac{\Gamma'\left(\nu\right)}{\left[\Gamma\left(\nu\right)\right]^{2}} &\to -1. \label{eq:2-4-4}
\end{align}
From Equation~(\ref{eq:2-4-2}) it follows that, 
\begin{align*}
    -\Gamma'\left(\nu\right)&=\Gamma\left(\nu\right)\left[\frac{1}{\nu}+\gamma-\Psi\left(1+\nu\right)-\gamma\right] \\
    &= \Gamma\left(\nu\right)\left[\frac{1}{\nu}-\Psi\left(1+\nu\right)\right], \hspace{1em} \nu>0.
\end{align*}
Therefore, 
\begin{align*}
    -\Gamma''\left(\nu\right)= \Gamma'\left(\nu\right)\left[\frac{1}{\nu}-\Psi\left(1+\nu\right)\right] + \Gamma\left(\nu\right)\left[-\frac{1}{\nu^{2}}-\Psi'\left(1+\nu\right)\right],
\end{align*}
and,
\begin{align*}
    \frac{-\Gamma''\left(\nu\right)}{\left[\Gamma\left(\nu\right)\right]^{3}} = \frac{\Gamma'\left(\nu\right)}{\left[\Gamma\left(\nu\right)\right]^{2}}\cdot \frac{1}{\Gamma\left(\nu\right)}\left[\frac{1}{\nu}-\Psi\left(1+\nu\right)\right]+\frac{1}{\left[\Gamma\left(\nu\right)\right]^{2}}\left[-\frac{1}{\nu^{2}}-\Psi'\left(1+\nu\right)\right].
\end{align*}
Remember that $\Psi'\left(1\right) = \frac{\pi^{2}}{6}$. By making use of Equation~(\ref{eq:2-4-4}) and the fact that $\nu\Gamma\left(\nu\right) \to 1$ as $\nu \to 0^{+}$, it follows that
\begin{align*}
    \frac{-\Gamma''\left(\nu\right)}{\left[\Gamma\left(\nu\right)\right]^{3}} \to\left(-1\right)\left(1-0\right)+\left(-1-0\right) = -2.
\end{align*}
Therefore, as $\nu \to 0^{+}$,
\begin{align}
    \frac{\Gamma''\left(\nu\right)}{\left[\Gamma\left(\nu\right)\right]^{3}} \to 2. \label{eq:2-4-5}
\end{align}
Substitute Equation~(\ref{eq:2-4-5}) into Equation~(\ref{eq:2-4-1})and assume that $n \geq 2$. Then,
\begin{align*}
    \lim_{\nu \to 0^{+}}\left\{\left[\frac{\Gamma''\left(\nu\right)}{\Gamma\left(\nu\right)\left[\Gamma\left(\nu\right)\right]^{2}}-\frac{\Gamma'\left(\nu\right)}{\left[\Gamma\left(\nu\right)\right]^{2}}\cdot\frac{\Gamma'\left(\nu\right)}{\left[\Gamma\left(\nu\right)\right]^{2}}\right]\cdot \frac{\left[\Gamma\left(\nu+\frac{1}{2}\right)\right]^{2n}}{\left[\Gamma\left(\nu\right)\right]^{2n-2}}\right\}^{\frac{1}{2}} = \left\{\left[2-\left(-1\right)^{2}\right]\cdot 0\right\}^{\frac{1}{2}} = 0,
\end{align*}
which follows from the fact that $\Gamma\left(\frac{1}{2}\right) = \sqrt{\pi}$ and $2n-2>0$, therefore $\frac{\left[\Gamma\left(\nu+\frac{1}{2}\right)\right]^{2n}}{\left[\Gamma\left(\nu\right)\right]^{2n-2}} \to \frac{\left(\sqrt{\pi}\right)^{2n}}{\infty} = 0$, if $n\geq 2$.
This means that $p_{1}\left(\nu|data\right) \to 0$ if $\nu \to 0^{+}$. A similar proof can be made for $p_{2}\left(\nu|data\right)$.
\end{proofEnd}

\section{Simulation Study} \label{sec:sims}
\subsection{Priors Compared} \label{ssec:simspriors}

The six priors that are used in the simulation study for comparison are:
\begin{enumerate}
    \item $p_{1}\left(\nu,\mu, \sigma^2\right) \propto \sigma^{-2} \left[\Psi'\left(\frac{\nu}{2}\right)-\Psi'\left(\frac{\nu+1}{2}\right)-\frac{2\left(\nu+5\right)}{\nu\left(\nu+1\right)\left(\nu+3\right)}\right]^{\frac{1}{2}}$
    \item $p_{2}\left(\nu,\mu, \sigma^2\right) \propto \sigma^{-2} \left[\Psi'\left(\frac{\nu}{2}\right)-\Psi'\left(\frac{\nu+1}{2}\right)-\frac{2\left(\nu+3\right)}{\nu\left(\nu+1\right)^{2}} \right]^{\frac{1}{2}}$
    \item $p_{3}\left(\nu,\mu, \sigma^2\right) \propto \sigma^{-2}\left(\frac{\nu}{\nu+3}\right)^{\frac{1}{2}} \left[\Psi'\left(\frac{\nu}{2}\right)-\Psi'\left(\frac{\nu+1}{2}\right)-\frac{2\left(\nu+3\right)}{\nu\left(\nu+1\right)^{2}} \right]^{\frac{1}{2}}$
    \item $p_{4}\left(\nu,\mu, \sigma^2\right) \propto \sigma^{-3}\left(\frac{\nu+1}{\nu+3}\right)^{\frac{1}{2}}\left(\frac{\nu}{\nu+3}\right)^{\frac{1}{2}} \left[\Psi'\left(\frac{\nu}{2}\right)-\Psi'\left(\frac{\nu+1}{2}\right)-\frac{2\left(\nu+3\right)}{\nu\left(\nu+1\right)^{2}} \right]^{\frac{1}{2}}$
    \item $p_{5}\left(\nu,\mu, \sigma^2\right) \propto \sigma^{-2}e^{-\xi\nu}$, where $\xi=0.1$
    \item $p_{6}\left(\nu,\mu, \sigma^2\right) \propto \sigma^{-2}\frac{2\nu d}{\left(\nu+d\right)^{3}}$, where $d=1.2$.
\end{enumerate}

For the standard univariate $t$-distribution ($\mu=0, \sigma^{2}=1$), the priors will be denoted by $p_{i}\left(\nu\right)$, $i=1,2,\ldots,6$. 

As mentioned in Appendix~\ref{app:proofsA}, $p_{1}\left(\nu,\mu, \sigma^2\right)$ is a reference prior with respect to the orderings $\{\mu,\sigma^2, \nu\}$, $\{\sigma^2, \mu, \nu\}$ and $\{\sigma^2, \nu, \mu\}$. From the Fisher information matrix (Equation~\ref{eq:fi}) it is clear that it is also a Jeffrey's prior for $\nu$ if $\mu$ and $\sigma^2$ are considered to be known. \citeasnoun{liu97} proposed the prior $\pi\left(\nu\right) \propto \left[\Psi'\left(\frac{\nu}{2}\right)-\Psi'\left(\frac{\nu+d}{2}\right)-\frac{2d\left(\nu+d+4\right)}{\nu\left(\nu+d\right)\left(\nu+d+2\right)} \right]^{\frac{1}{2}}$ for the multivariate $t$-distribution, where $d$ is the dimension of the multivariate distribution. The prior $\pi\left(\nu\right)$ is obtained by applying Jeffreys' rule \cite{box11}. \citeasnoun{villa18} included it in their simulation study on objective priors for the number of degrees of freedom of a multivariate $t$-distribution. If $d=1$, $\pi\left(\nu\right)$ simplifies to $p_{1}\left(\nu\right)$. 

The prior $p_{2}\left(\nu,\mu, \sigma^2\right)$ on the other hand is a probability-matching prior for $\nu$ as well as a reference prior for the parameter orderings $\{\nu,\mu, \sigma^2\}$, $\{\mu,\nu, \sigma^2\}$ and $\{\nu,\sigma^2, \mu\}$ (see Theorems~\ref{theorem:ref_prior} and \ref{theorem:pm_prior}).

The prior $p_{3}\left(\nu,\mu, \sigma^2\right)$ is the independence Jeffreys prior and $p_{4}\left(\nu,\mu, \sigma^2\right)$ is the Jeffreys-rule prior. Both of these priors were derived by \citeasnoun{fonseca08}. The Jeffreys-rue prior is proportional to the square root of the determinant of the Fisher information matrix while the independence Jeffreys prior is obtained by assuming that the priors for $\mu$ and $\left(\sigma^{2},\nu\right)$ are independent, \textit{i.e.}  $p_{3}\left(\nu,\mu, \sigma^2\right)= p_{3}\left(\mu\right) p_{3}\left(\nu, \sigma^2\right)$. From the Fisher information matrix defined in Equation~\ref{eq:fi} it therefore follows that  $$p_{3}\left(\mu\right)\propto \sqrt{\det\left[I\left(\theta\right)\right]_{22}} \propto 1$$ and $$p_{3}\left(\nu, \sigma^2\right)\propto \sqrt{\left[I\left(\theta\right)\right]_{11}\left[I\left(\theta\right)\right]_{33}-\left[I\left(\theta\right)\right]_{13}^{2}}.$$

The exponential prior $p_{5}\left(\nu,\mu, \sigma^2\right) $ was derived by \citeasnoun{geweke93} but according to \citeasnoun{fonseca08} and \citeasnoun{villa14} this prior is too informative and is found to dominate the data. 

\citeasnoun{juarez10} considered a non-hierarchical and a hierarchical prior. The first is a gamma prior with parameters 1 and $1/100$. The hierarchical prior is obtained by considering an exponential distribution for the scale parameter of the gamma prior with slope parameter $a$. In other words, $$p_{6}\left(\nu\right) = \int_{0}^{\infty}p\left(\nu|a\right)p\left(a|\tilde{d}\right) da, $$ where $p\left(\nu|a\right)=a^{2}\nu e^{-a\nu}$ and $p\left(a|\tilde{d}\right)=\tilde{d}e^{-a\tilde{d}}$. The resulting prior is therefore $p_{6}\left(\nu,\mu, \sigma^2\right) \propto \sigma^{-2}\frac{2\nu \tilde{d}}{\left(\nu+\tilde{d}\right)^{3}}$ for $\nu>0$ and $\tilde{d}>0$. The parameter $a$ controls the mode $\left(\tilde{d}/2\right)$ and the median $\left(1+\sqrt{2}\right)\tilde{d}$. \citeasnoun{rubio15} mention that if $\tilde{d}=1.2$ then it will be a continuous alternative to the discrete objective prior proposed by \citeasnoun{villa14}.

\subsection{Frequentist Properties} \label{ssec:simsprop}
In this subsection we summarize the frequentist properties of the priors for $\nu$ in the case of the univariate standard $t$-distribution. The focus is on the relative square rooted mean square error from the median of the posterior distribution of $\nu$. The index is denoted by $\sqrt{MSE\left(\nu\right)}/\nu$ where $MSE=E\left(\nu-m\right)^{2}$. The frequentist coverage percentages of the 95\% credibility intervals for a sample of size $n=30$ as well as the interval lengths and modes of the interval lengths based on 2000 simulations are also considered.  

\begin{table}
\caption{Relative Root Mean Squared Errors ($\sqrt{MSE\left(\nu\right)}/\nu$) for Six Priors for $\nu$}
\centering
\begin{tabular}[ht]{|l||r|r|r|r|r|r|} \hline \label{tab:rrmse}
$\nu$&\textbf{Prior 1}&\textbf{Prior 2}&\textbf{Prior 3}&\textbf{Prior 4}&\textbf{Prior 5}&\textbf{Prior 6}\\ \hline \hline
1&0.4593&0.4138&0.4707&0.4416&0.5703&0.6153\\ \hline
2&0.8714&0.8408&0.9546&0.9650&1.1661&1.1180\\ \hline
3&0.9264&0.9120&1.0001&1.0280&1.1807&1.1226\\ \hline
4&0.7946&0.7779&0.8377&0.8840&0.9811&1.0013\\ \hline
5&0.6995&0.6991&0.7560&0.7768&0.8525&0.9068\\ \hline
6&0.6202&0.5825&0.6501&0.6563&0.7110&0.7649\\ \hline
7&0.5062&0.5014&0.5335&0.5565&0.5705&0.6498\\ \hline
8&0.4416&0.4316&0.4495&0.4805&0.4620&0.5320\\ \hline
9&0.4055&0.3975&0.4058&0.4268&0.3916&0.4674\\ \hline
10&0.3899&0.3850&0.3871&0.3928&0.3519&0.4180\\ \hline
11&0.4022&0.3961&0.3905&0.3860&0.3367&0.3893\\ \hline
12&0.3982&0.3909&0.3841&0.3802&0.3214&0.3760\\ \hline
13&0.4098&0.4138&0.4069&0.3961&0.3378&0.3698\\ \hline
14&0.4174&0.4341&0.4181&0.4053&0.3510&0.3648\\ \hline
15&0.4385&0.4528&0.4327&0.4242&0.3685&0.3797\\ \hline
16&0.4583&0.4703&0.4508&0.4332&0.3873&0.3942\\ \hline
17&0.4789&0.4918&0.4724&0.4573&0.4143&0.4160\\ \hline
18&0.5107&0.5155&0.5050&0.4819&0.4486&0.4425\\ \hline
19&0.5178&0.5291&0.5021&0.4953&0.4534&0.4510\\ \hline
20&0.5403&0.5473&0.5353&0.5217&0.4875&0.4761\\ \hline
21&0.5581&0.5662&0.5479&0.5359&0.5023&0.4871\\ \hline
22&0.5646&0.5778&0.5541&0.5474&0.5133&0.5032\\ \hline
23&0.5834&0.5896&0.5725&0.5626&0.5327&0.5204\\ \hline
24&0.5978&0.6101&0.5902&0.5764&0.5541&0.5386\\ \hline
25&0.6064&0.6153&0.5921&0.5816&0.5582&0.5418\\ \hline \hline
\textbf{Mean}&\textbf{0.5439}&\textbf{0.5417}&\textbf{0.5520}&\textbf{0.5517}&\textbf{0.5522}&\textbf{0.5699}\\ \hline
\end{tabular}
\end{table}

From Table~\ref{tab:rrmse} it is clear that the reference priors $p_{1}\left(\nu\right)$ and $p_{2}\left(\nu\right)$ are on average the two best priors. The prior $p_{2}\left(\nu\right)$, which is also a probability-matching prior, is somewhat better than $p_{1}\left(\nu\right)$. The prior $p_{6}\left(\nu\right)$ is performing worst in this study. 

\begin{table} 
\caption{Averages of the Relative Root Mean Squared Errors for $\nu$}
\centering
\begin{tabular}[ht]{|l||r|r|r|r|r|r|} \hline \label{tab:avgrrmse}
$\nu$&\textbf{Prior 1}&\textbf{Prior 2}&\textbf{Prior 3}&\textbf{Prior 4}&\textbf{Prior 5}&\textbf{Prior 6}\\ \hline \hline
\textbf{Mean (1 to 10)}&0.6115&0.5942&0.6445&0.6608&0.7238&0.7596\\ \hline
\textbf{Mean (11 to 25)}&0.4988&0.5067&0.4903&0.4790&0.4378&0.4434\\ \hline
\end{tabular}
\end{table}

From Table~\ref{tab:avgrrmse} it can be seen that $p_{2}\left(\nu\right)$ is particularly good if $\nu$ is small (1 to 10).  Researchers are usually interested in $t$-distributions with a small number of degrees of freedom. For large values of $\nu$ (11 to 25) we have that $p_{5}\left(\nu\right)$ and $p_{6}\left(\nu\right)$ seem to be the best priors. 

\begin{figure}[h]
    \centering
    \caption{Relative Root Mean Squared Errors for the Posterior Median of $\nu$}
    \vspace{-1em}
    \includegraphics[width=.8\linewidth]{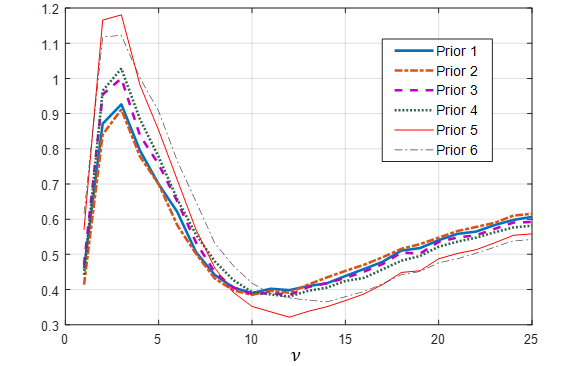}
    \label{fig:rrmse}
\end{figure}

It is clear from Figure~\ref{fig:rrmse} that for $\nu$ between two and six the $\sqrt{MSE(\nu)}/\nu$ values for priors $p_{5}\left(\nu\right)$ and $p_{6}\left(\nu\right)$ are larger than those of the objective priors. 

\begin{table} 
\caption{Coverage Percentages of the 95\% Credibility Intervals for $\nu$}
\centering
\begin{tabular}[ht]{|l||r|r|r|r|r|r|} \hline \label{tab:cov}
$\nu$&\textbf{Prior 1}&\textbf{Prior 2}&\textbf{Prior 3}&\textbf{Prior 4}&\textbf{Prior 5}&\textbf{Prior 6}\\ \hline \hline
1&94.10&94.50&94.20&93.90&91.15&92.75\\ \hline
2&94.25&94.80&93.80&93.95&86.55&92.00\\ \hline
3&97.55&97.15&97.20&96.45&88.65&96.70\\ \hline
4&97.75&97.70&98.60&97.70&97.70&98.80\\ \hline
5&97.10&97.20&97.30&98.05&99.30&98.40\\ \hline
6&97.30&97.15&97.60&98.00&99.40&98.25\\ \hline
7&97.90&97.55&98.20&97.95&99.75&98.55\\ \hline
8&97.30&97.60&97.80&97.75&99.75&98.30\\ \hline
9&98.35&97.65&98.20&97.75&99.30&98.60\\ \hline
10&97.70&97.05&97.80&98.25&98.85&98.20\\ \hline
11&96.90&96.60&97.20&98.30&98.90&97.75\\ \hline
12&97.60&97.15&97.90&98.45&99.75&97.95\\ \hline
13&97.55&97.45&97.80&97.85&99.65&98.55\\ \hline
14&97.80&97.45&98.10&98.10&99.35&98.65\\ \hline
15&97.85&96.80&97.60&98.15&99.10&98.45\\ \hline
16&97.85&97.10&98.50&98.60&99.10&98.35\\ \hline
17&97.75&97.95&98.20&97.85&99.30&98.65\\ \hline
18&97.10&96.30&97.00&97.80&98.90&98.30\\ \hline
19&97.65&97.30&98.00&97.70&99.00&98.25\\ \hline
20&96.80&97.00&96.60&97.80&98.50&97.40\\ \hline
21&95.90&96.30&96.50&97.70&98.10&97.50\\ \hline
22&97.00&96.10&97.50&97.60&98.80&98.15\\ \hline
23&97.15&97.10&98.00&97.05&98.70&97.85\\ \hline
24&96.55&95.55&96.70&98.45&97.40&97.75\\ \hline
25&97.20&97.10&98.40&97.35&99.00&98.60\\ \hline \hline
\textbf{Mean}&\textbf{97.118}&\textbf{96.864}&\textbf{97.388}&\textbf{97.540}&9\textbf{7.758}&\textbf{97.708}\\ \hline
\end{tabular}
\end{table}

\begin{figure}[h]
    \centering
    \caption{Frequentist Coverage Percentages of the 95\% Credibility Intervals for $\nu$}
    \vspace{-1em}
    \includegraphics[width=.8\linewidth]{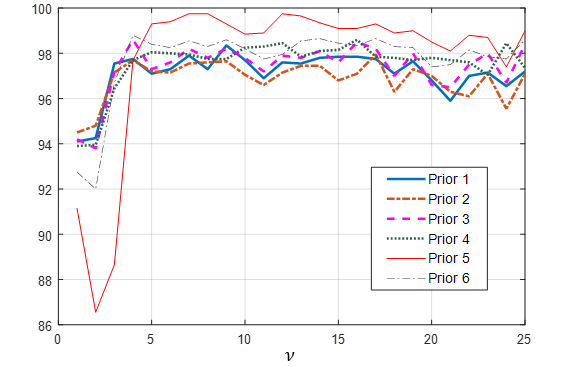}
    \label{fig:cov}
\end{figure}

In Table~\ref{tab:cov} the frequentist coverage percentages of the 95\% credibility intervals are given for a sample size of $n=30$ and 2000 simulations. In Figure~\ref{fig:cov} these intervals are illustrated graphically. 

According to \citeasnoun[Figure 2, p. 329]{fonseca08} the coverage percentages of the 95\% credibility intervals for $\nu$ in the case of the Jeffrey's-rule prior, are poor for $n=30$. They also mentioned that for the Geweke prior, $p_{5}\left(\nu\right)$, the frequentist coverage is much smaller than the nominal level for small $\nu$ and is undesirably close to 1 for $\nu > 6$. 

The results of \citeasnoun{fonseca08} differ somewhat from our results given in Table~\ref{tab:cov} and Figure~\ref{fig:cov}. From Table~\ref{tab:cov} it can be seen that for $\nu \leq 3$ the frequentist coverage percentage of the Geweke prior is smaller than the nominal level and for $\nu \geq 4$ it is on average 98.98\%. The coverage percentages of the Jeffreys-rule prior $p_{4}\left(\nu\right)$, however, do not differ much from those of the other objective priors ($p_{1}\left(\nu\right)$, $p_{2}\left(\nu\right)$ and $p_{3}\left(\nu\right)$). In the case of the coverage percentages, the reference (or probability-matching) prior $p_{2}\left(\nu\right)$ seems to be the best because it has on average a 96.86\% coverage. 

\begin{table} 
\caption{Average Interval Lengths of the 95\% Credibility Intervals for $\nu$}
\centering
\begin{tabular}[t]{|l||r|r|r|r|r|r|} \hline \label{tab:avg_length}
$\nu$&\textbf{Prior 1}&\textbf{Prior 2}&\textbf{Prior 3}&\textbf{Prior 4}&\textbf{Prior 5}&\textbf{Prior 6}\\ \hline \hline
1&1.4863&1.4705&1.5211&1.6613&1.6616&1.7346\\ \hline
2&10.9029&10.2188&11.4415&11.4660&8.7877&13.1349\\ \hline
3&27.2776&25.6413&28.6591&27.4616&16.8940&31.0931\\ \hline
4&40.8181&39.1123&42.8603&46.0271&22.6912&49.8889\\ \hline
5&58.0892&54.0122&60.7860&57.8947&27.6600&70.0064\\ \hline
6&66.8981&64.0113&69.8454&70.4983&29.9378&80.6659\\ \hline
7&74.5588&72.6509&77.8293&78.6948&32.0676&91.6501\\ \hline
8&81.3848&78.6516&84.8817&88.3929&33.6076&98.8977\\ \hline
9&84.4323&80.7490&87.9760&94.3561&34.2755&105.0209\\ \hline
10&90.9408&87.5548&94.5730&96.4120&35.2627&109.2556\\ \hline
11&91.3351&87.6219&94.9579&104.1987&35.2547&110.6545\\ \hline
12&97.0578&93.8699&100.7701&103.0313&36.2704&114.0066\\ \hline
13&96.1922&93.0008&99.8689&105.4707&36.1493&117.2605\\ \hline
14&99.1662&95.2863&102.9146&110.7050&36.6309&122.5355\\ \hline
15&101.8962&97.7551&105.7096&108.7843&37.0405&124.0829\\ \hline
16&103.5499&100.0941&107.4031&115.3113&37.3834&126.0403\\ \hline
17&104.6025&100.4079&108.4889&114.0209&37.5741&126.7225\\ \hline
18&102.7811&99.5750&106.6120&115.6712&37.2385&126.3335\\ \hline
19&108.6713&104.4110&112.4989&115.4082&38.0129&129.7908\\ \hline
20&105.8102&102.0438&109.6180&115.2661&37.5957&129.2041\\ \hline
21&107.9256&104.7400&111.7811&115.8514&37.8954&131.9269\\ \hline
22&110.7806&106.7750&114.5896&116.6591&38.3138&132.9114\\ \hline
23&111.1055&107.3003&114.9879&117.8916&38.4163&133.3977\\ \hline
24&110.3250&105.5900&114.1435&120.3860&38.2195&133.3991\\ \hline
25&114.3742&110.4022&118.2419&122.1614&38.8475&137.3235\\ \hline \hline
\textbf{Mean}&\textbf{84.0945}&\textbf{80.9178}&\textbf{87.3184}&\textbf{90.9473}&\textbf{32.1475}&\textbf{101.8775}\\ \hline
\end{tabular}
\end{table}

\begin{figure}[h]
    \centering
    \caption{Average Interval Lengths of the 95\% Credibility Intervals for $\nu$}
    \vspace{-1em}
    \includegraphics[width=.8\linewidth]{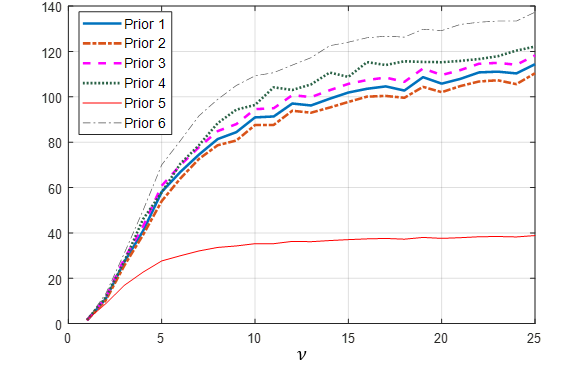}
    \label{fig:avg_length}
\end{figure}

From Table~\ref{tab:avg_length} and Figure~\ref{fig:avg_length} it can be observed that $p_{2}\left(\nu\right)$ has the shortest average interval lengths of all the objective priors. The prior that gives the shortest interval lengths is however $p_{5}\left(\nu\right)$, the Geweke prior, with interval lengths on average two and a half to three times shorter than those of the objective priors and with a coverage percentage of more than 95\%. The worst performing prior seems to be $p_{6}\left(\nu\right)$.

\begin{figure}[h!]
    \centering
    \caption{Mode of Interval Lengths of the 95\% Credibility Intervals for $\nu$}
    \vspace{-1em}
    \includegraphics[width=.8\linewidth]{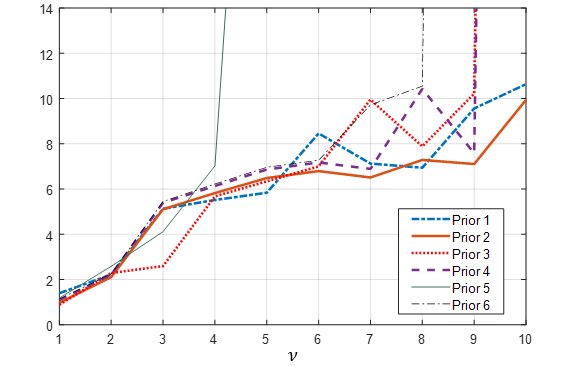}
    \label{fig:mode_lengths}
\end{figure}

Although the interval lengths of the objective priors for most of the 2000 simulations are quite small, a few extremely large lengths can have a big influence on the average interval length. A large interval length will occur if the observations in the sample are of such a nature that it is not clear if the data were drawn from a normal or $t$-distribution. It is for this reason that the modes of the interval lengths are given in Table~\ref{tab:mode_lengths} and Figure~\ref{fig:mode_lengths}.

\begin{table} 
\caption{Mode of Interval Lengths of the 95\% Credibility Intervals for $\nu$}
\centering
\begin{tabular}[t]{|l|l|l|l|l|l|l|} \hline \label{tab:mode_lengths}
$\nu$&\textbf{Prior 1}&\textbf{Prior 2}&\textbf{Prior 3}&\textbf{Prior 4}&\textbf{Prior 5}&\textbf{Prior 6}\\ \hline \hline
1&1.3918&0.9694&0.8824&1.0916&1.1767&1.1350\\ \hline
2&2.2132&2.1039&2.2849&2.2145&2.5818&2.2474\\ \hline
3&5.1277&5.1071&2.5923&5.3916&4.1166&5.4524\\ \hline
4&5.5141&5.8186&5.6744&6.1332&7.0209&6.2217\\ \hline
5&5.8345&6.4857&6.3385&6.8614&40.3448&6.9601\\ \hline
6&8.4593&6.7914&7.0016&7.1734&39.8707&7.2716\\ \hline
7&7.1246&6.5109&9.9536&6.8896&42.4445&9.7214\\ \hline
8&6.9421&7.2842&7.8857&10.4286&40.2139&10.5493\\ \hline
9&9.5604&7.1040&10.2161&7.5794&41.2099&162.6129\\ \hline
10&10.6371&9.9450&151.4783&156.0163&40.6707&178.7038\\ \hline
\end{tabular}
\end{table}

As before, the reference priors $p_{1}\left(\nu\right)$ and $p_{2}\left(\nu\right)$ seem to be the two best priors because the modes of their interval lengths are in general the smallest. The prior $p_{2}\left(\nu\right)$ seems to be somewhat better than $p_{1}\left(\nu\right)$. The modes of the interval lengths of the priors $p_{3}\left(\nu\right)$ and $p_{4}\left(\nu\right)$ (the independence Jeffreys and the Jeffreys-rule priors) change dramatically for $\nu \geq 10$. From Table~\ref{tab:mode_lengths} it is clear that $p_{5}\left(\nu\right)$, the Geweke prior, is the worst prior for $5 \leq \nu \leq 8$. It does well for $1 \leq \nu \leq 4$ and seems to do better than most of the priors for $\nu > 10$. The prior $p_{6}\left(\nu\right)$ again seems to perform worst in this study.

\section{Application} \label{sec:appl}
To compare the six priors on real data, a random sample of $n=100$ observations of the daily log-returns of IBM data is analysed. The original data set contains 2528 observations for the period from the $3^{\textnormal{rd}}$ of January 1989 to the $31^{\textnormal{st}}$ of December 1998. The data are available from the `Ecdat' R package \cite{Rcoreteam,Croissant2015}. 

\begin{figure}[ht!]
    \centering
    \captionsetup{width=.75\linewidth}
    \caption{Posterior Distribution of $\mu$ using Prior $p_{2}\left(\nu,\mu, \sigma^2\right)$}
    \vspace{-1em}
    \includegraphics[width=.8\linewidth]{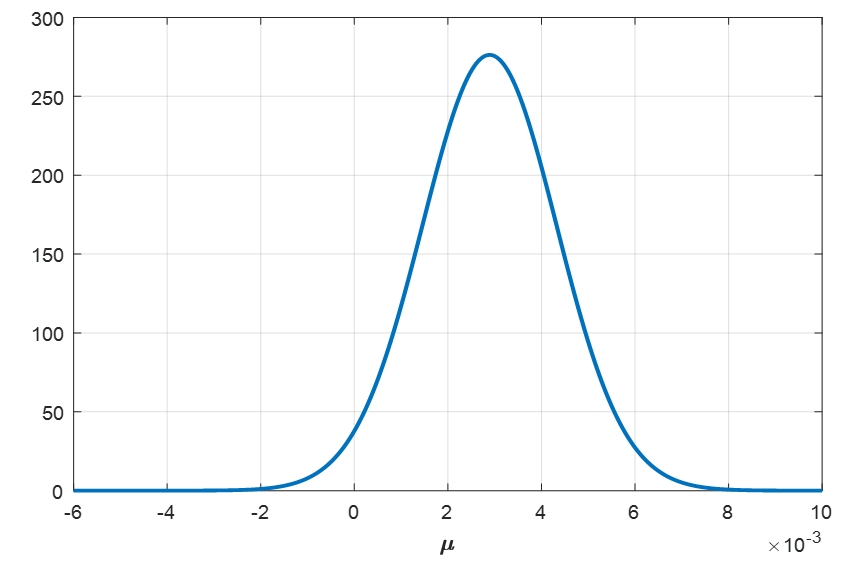}
    \label{fig:post1}
    \caption*{Mean $=0.0029$, Median $=0.0029$, Mode $=0.00289$, Var $=2.0970\times10^{-6}$; 95\% Equal-tail Interval $=\left(2.12\times 10^{-5}; 0.0057\right)$}
\end{figure}

\begin{figure}[h!]
    \centering
    \captionsetup{width=.75\linewidth}
    \caption{Posterior Distribution of $\sigma^{2}$ using Prior $p_{2}\left(\nu,\mu, \sigma^2\right)$}
    \vspace{-1em}
    \includegraphics[width=.8\linewidth]{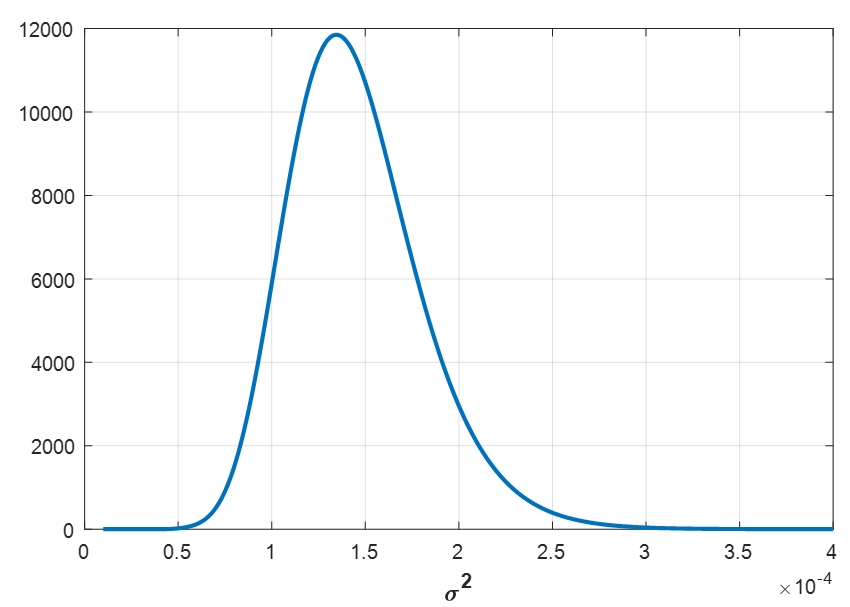}
    \label{fig:post2}
    \caption*{Mean $=1.4373\times 10^{-4}$, Median $=1.4006\times10^{-4}$, Mode $=1.35\times10^{-4}$, Var $=1.2229\times10^{-9}$; 95\% Equal-tail Interval $=\left(8.548\times 10^{-5}; 2.252\times 10^{-4}\right)$}
\end{figure}

\begin{figure}[h!]
    \centering    
    \captionsetup{width=.75\linewidth}
    \caption{Posterior Distribution of $\nu$ using Prior $p_{2}\left(\nu,\mu, \sigma^2\right)$}
    \vspace{-1em}
    \includegraphics[width=.8\linewidth]{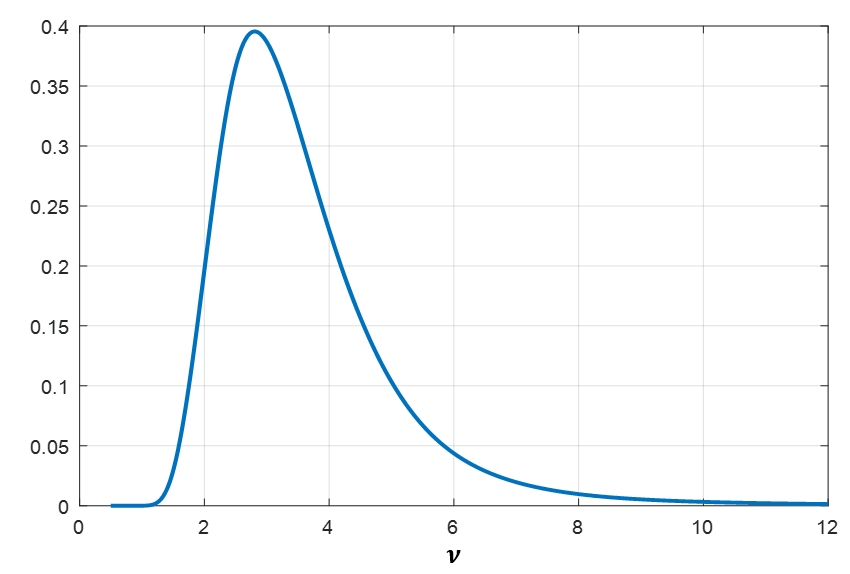}
    \label{fig:post3}
    \caption*{Mean $=3.6118$, Median $=3.28$, Mode $=2.810$, Var $=2.2258$; 95\% Equal-tail Interval $=\left(1.797; 7.136\right)$}
\end{figure}

By using the prior $p_{2}\left(\nu,\mu, \sigma^2\right) \propto \sigma^{-2} \left[\Psi'\left(\frac{\nu}{2}\right)-\Psi'\left(\frac{\nu+1}{2}\right)-\frac{2\left(\nu+3\right)}{\nu\left(\nu+1\right)^{2}} \right]^{\frac{1}{2}}$ and Gibbs sampling the posterior distribution of the parameters $\mu, \sigma^{2}$ and $\nu$ are obtained and illustrated in Figures~\ref{fig:post1}, \ref{fig:post2} and \ref{fig:post3}.  The resulting posterior statistics of $\nu$ for the six priors are summarized in Table~\ref{tab:post_stats}. The conditional posterior distributions that were used in the Gibbs sampling procedure are given in Appendix~\ref{app:proofsC}.

\begin{table}{h!}
\centering
\caption{Posterior Statistics obtained by using Six Different Priors for $\nu$}
\begin{tabular}[t]{|l||r|r|r|} \hline \label{tab:post_stats}
Prior&Mean&Median&95\% Credibility Interval\\ \hline \hline
1&3.6454&3.33&(1.851; 7.044)\\ \hline
2&3.6118&3.28&(1.797; 7.136)\\ \hline
3&3.7471&3.41&(1.830; 7.217)\\ \hline
4&3.7894&4.43&(1.860; 7.919)\\ \hline
5&4.3722&3.95&(2.064; 9.112)\\ \hline
6&3.9485&3.59&(1.940; 7.561)\\ \hline
\end{tabular}
\end{table}

It can be noticed that the posterior statistics of $\nu$ for the objective priors are for all practical purposes the same, but differ somewhat from those of the exponential ($p_{5}\left(\nu,\mu, \sigma^2\right)$) and the hierarchical ($p_{6}\left(\nu,\mu, \sigma^2\right)$) priors.

\section{Discussion} \label{sec:disc}
The Student $t$-distribution is of great importance for many economic and business data because is reduces the influence of outliers in model estimation and thus makes statistical analysis more robust. 

Unfortunately, the estimation of $\nu$, the number of degrees of freedom of the $t$-distribution is not easy. The reason for this is the bad behaviour of the likelihood function for $\nu$. To overcome the fact that the likelihood function does not vanish in the tail a prior distribution that tends to zero as $\nu$ tends to infinity should be applied. It is for this reason that two non-informative priors, $p_{1}\left(\nu,\mu, \sigma^2\right)$ and $p_{2}\left(\nu,\mu, \sigma^2\right)$, have been derived for the parameters of the Student $t$ distribution. Both of these priors are reference priors while $p_{2}\left(\nu,\mu, \sigma^2\right)$  is also a probability-matching prior. 

Our simulation studies illustrate the good frequentist properties of the posterior distributions associated to these priors. The focus has been on the relative square-rooted mean squared error from the posterior median and the 95\% credibility intervals for a sample of size $n=30$ based on 2000 simulations. We have compared the frequentist properties of the two reference priors to four other priors (the Jeffrey's-rule prior, the independence Jeffreys prior, the exponential prior and a hierarchical prior). Overall the two reference priors seem to give better results, especially if $1\leq \nu \leq 10$.

In Section~\ref{sec:appl} the six priors are compared on a real data set. A random sample of $n=100$ observations of the daily log-returns of IBM data is analysed. The results show that the posterior statistics of $\nu$ for the objective priors are for all practical purposes the same, but differ somewhat from those of the exponential and hierarchical priors. 

\bibliography{references}

\begin{appendices}

\section{} \label{app:proofsA}
This appendix provides derivations for the reference and probability-matching priors for the univariate Student $t$-distribution. As in the case of the Jeffreys' priors the derivations of these priors are based on the Fisher information matrix. Differentiation of the log likelihood functions twice with respect to the unknown parameters and taking expected values gives the Fisher information matrix. 

\begin{align*}
\left\{I\left(\theta\right)\right\}_{ij} & = E_{X|\theta}\left[-\frac{\partial^{2}}{\partial\theta_{i} \partial\theta_{j}} \log \left\{L\left(\btheta;\bx\right)\right\} \right],
\end{align*}
where $\btheta = \left[\begin{array}{ccc} \theta_{1} & \theta_{2} & \theta_{3} \end{array}\right]' = \left[\begin{array}{ccc} \nu & \mu & \sigma \end{array}\right]$, and,
\begin{align*}
    L\left(\theta,\bx\right) &= \prod_{i=1}^{n}f\left(x_{i}|\mu,\sigma^{2},\nu\right) \\
    &=\frac{\left[\Gamma\left(\frac{\nu+1}{2}\right)\right]^{n} \nu^{n\nu/2}}{\left[\Gamma\left(\frac{\nu}{2}\right)\right]^{n} \left[\Gamma\left(\frac{1}{2}\right)\right]^{n} \sigma^{n}} \prod_{i=1}^{n} \left[ \nu + \left(\frac{x_{i}-\mu}{\sigma}\right)^{2} \right]^{-\frac{1}{2}\left(\nu+1\right)}.
\end{align*}

\printProofs[A]

\section{} \label{app:proofsB}

\printProofs[B]

\section{} \label{app:proofsC}

If $x_{i}|\mu,\sigma^{2},\lambda_{i} \sim N\left(\mu, \frac{\sigma^{2}}{\lambda_{i}}\right), i=1,2,\ldots,n$, and $\nu \lambda_{i}\sim \chi^{2}_{\nu}$, then $x_{i}|\mu,\sigma^{2},\nu \sim t_{\nu}\left(\mu,\sigma^{2}\right)$. If the prior $p_{j}\left(\mu,\sigma^{2},\nu\right) \propto \sigma^{-2}p_{j}\left(\nu\right), j=1,2,\ldots,6$ is used, then the following conditional posterior distributions can be derived:

\begin{align}
    \mu | \sigma^{2},H,\bx \sim N\left[\left(\bone' H \bone \right)^{-1} \bone' H \bx, \sigma^{2}\left(\bone' H \bone \right)^{-1} \right], \label{eq:postmu}
\end{align}
where $\bone = \left[\begin{array}{cccc} 1 & 1 & \cdots & 1 \end{array}\right]'$, $\bx = \left[\begin{array}{cccc} x_{1} & x_{2} & \cdots & x_{n} \end{array}\right]'$, and $H = \textnormal{diag}\left[\begin{array}{cccc} \lambda_{1} & \lambda_{2} & \cdots & \lambda_{n} \end{array}\right]' $;
\begin{align}
    \sigma^{2}|\mu,H,\bx \sim \frac{\left(\bx-\mu \bone\right)'H \left(\bx-\mu \bone\right)}{\chi^{2}_{n}}; \label{eq:postsig2}
\end{align}
\begin{align}
    \lambda_{i}|\mu, \sigma^{2},\nu,x_{i} \sim \frac{\chi^{2}_{\nu+1}}{\nu+\left(\frac{x_{i}-\mu}{\sigma^{2}}\right)^{2}}, i = 1,2,\ldots,n;\label{eq:postlambda}
\end{align}
and,
\begin{align}
    p\left(\nu|\lambda_{1},\lambda_{2},\ldots,\lambda_{n}\right) \propto \frac{\nu^{n\nu / 2}}{2^{n \nu /2}\left[\Gamma\left(\frac{\nu}{2}\right)\right]^{n}}\prod_{i=1}^{n} \left[\lambda_{i}^{\frac{1}{2}\left(\nu-1\right)} e^{-\frac{\nu}{2}\sum_{i=1}^{n}\lambda_{i}} \right]p_{j}\left(\nu\right) \label{eq:postnu}
\end{align}
By using Equations~(\ref{eq:postmu}), (\ref{eq:postsig2}), (\ref{eq:postlambda}) and (\ref{eq:postnu}) and then Gibbs sampling, the unconditional posterior distributions of $\mu$, $\sigma^{2}$ and $\nu$ can be obtained.  

In the case of $p_{4}\left(\mu,\sigma^{2},\nu\right)\propto \sigma^{-3}p_{4}\left(\nu\right)$, the degrees of freedom of the Chi-square distribution in Equation~(\ref{eq:postsig2}) changes to $n+1$.
\end{appendices}

\end{document}